\documentstyle[aps,preprint]{revtex}

\begin{document}

\draft

\title{
Detection of pairing from the extended Aharonov-Bohm period
in strongly correlated electron systems
}

\author{K. Kusakabe}
\address{
Institute for Solid State Physics, University of Tokyo, Roppongi,
Tokyo 106, Japan
}

\author{H. Aoki}
\address{
Department of Physics, University of Tokyo, Hongo, Tokyo 113, Japan
}

\maketitle

\begin{abstract}
Inspired from Sutherland's work [Phys. Rev. Lett. {\bf 74}, 816 (1995)]
on detecting bound spin waves,
we propose that bound electron states can be
detected from the dependence of interacting electron systems to
the Aharonov-Bohm flux in the `extended zone' scheme,
where the electron pairing halves the original period $N_a$
flux quanta in a system of linear size $N_a$.
Along with the Bethe-ansatz analysis, a numerical implementation
for keeping track of the adiabatic flow of energy levels
is applied to the attractive/repulsive Hubbard models and the $t-J$ ladder.
\end{abstract}
\vspace*{0.5cm}

\narrowtext

The response to an adiabatic change against external parameters
is an interesting way to probe the nature of interacting electron systems.
Some thirty years ago, Byers and Yang\cite{byers} proposed a notable
example of detecting the Cooper pairing:
while a normal state responds to
an Aharonov-Bohm (AB) flux periodically with the period
of flux quantum, $\Phi_0 \equiv hc/e$, a BCS state will have
a halved period, $\Phi_0 /2$.
The anomalous flux quantization has actually been
applied to various strongly correlated electron systems that
are intended to describe
high-$T_c$ cuprates\cite{fye,ferretti,dagotto,assaad}.

This is in fact one out of several ways to detect
superconductivity in purely electronic systems.
Since an effective electron-electron attraction {\it per se}
does not guarantee the Cooper pairing,
a usual way is to search for a long-tailed pairing correlation function,
but the finite-size effect must be carefully analyzed.
The quantum Monte Carlo method
(QMC)\cite{sorella,moreo,imada},
the density-matrix renormalization group\cite{white},
evaluation of the superfluid density
(helicity modulus)\cite{scalapino,dagotto}
are toward this line of approach.
Thus the detection of the Cooper pairing
(or bound electrons in more general terms) is a demanding problem
in a correlated electron gas,
except for one-dimensional(1D) systems, where
exact analytic treatment is feasible
with the Bethe-ansatz solutions coupled with the conformal field theory.

This is where the Byers-Yang flux quantization comes in.
The test, however, remains some way from a clear-cut criterion,
since the flux dependence of the ground-state energy
may give the half-flux periodicity even for
the repulsive Hubbard model\cite{ferretti}.
An origin of the obscured period is the spin degrees of
freedom\cite{kusmar}, which is most prominent in 1D.

In the present paper, we propose a new way to detect
bound electron states from a more global look at the response to the flux.
The idea has been inspired by a recent analysis of the bound spin waves
by Sutherland\cite{sutherland}:
bound complexes of spin waves
in 1D Heisenberg magnets
(or equivalently a gas of charged bosons)
may be detected from their response to a boosted total momentum.
The momentum boost is achieved by
twisting the boundary condition,
$\Psi (\ldots ,x_j+N_a,\ldots )=e^{i2\pi \phi}
\Psi (\ldots ,x_j,\ldots )$, for an $N_a$-site lattice,
which uniformly shifts the set of $k$ points.
He has shown that,
if all of the $N$ particles ({\it i.e.}, $N$ flipped spins in a magnet)
form one bound state, the energy returns
to its initial value by a twist of $\phi =N_a/N$, which is
$1/N$ times the twist $\phi = N_a$ required
to shift the set of $k$ points for the free particles
back to the original position.
Intuitively, this discerns whether
the momentum boost acts on individual particles
or on a `center-of-mass' of bound particles.
A key in Sutherland's idea is
to keep track of the $\phi$-dependence of the state
not over the one period  ($0\leq \phi <1$) but over
the `extended zone' ($0\leq \phi <N_a$).

Sutherland does not argue what happens
in {\it electron} systems or in the situation where more than
one bound complexes exist,
but we conjecture here that the extended AB method should,
conceptually, hold in general.
We would then be able to treat {\it e.g.} the Cooper pairing problem.
Natural questions are:
(i) Can we really extend Sutherland's spin-wave analysis to
electron systems?
(ii) Can we apply the method to two or higher dimensions?

In the present Letter
we first give a straightforward extension of Sutherland's spin analysis
to electron systems with the Bethe-ansatz analysis of the 1D
Hubbard model as a prototype.
There we introduce an AB flux $\Phi$ that couples to the charge
degrees of freedom to twist the boundary condition.
We then look at the energy levels against $\Phi$ over
$0\leq \Phi <N_a\Phi_0$ (which we call the `extended AB' spectral flow)
to discriminate the
bound states,
as opposed to the conventional
wisdom that one period of $\Phi_0$ suffices.
We next propose a method to numerically
implement the extended AB test for arbitrary systems including
2D systems to go beyond the Bethe-ansatz analysis.

We start with confirming for the 1D Hubbard model that
the electrons have a reduced period of $N_a\Phi_0/N$ as well
for $N$-bound states with
the Bethe-ansatz analysis\cite{lw}.
Consider an $N_a$-site ring
(with $N_a$ even for simplicity)
containing $N$ electrons, and thread a magnetic flux $\Phi$.
A change of $\Phi$ by $\Phi_0$ shifts the
$k$ points exactly by their spacing $\Delta k=2\pi/N_a$
for noninteracting electrons,
so that the set of $k$ points
accomplishes a full travel across the Brillouin zone
when $\Phi/\Phi_0$ reaches $N_a$.
When interacting, the electrons are subject to
Bethe-ansatz equations\cite{ss1},
for which there are two types of solutions, i.e.,
real and complex roots\cite{takahashi}.
A real charge rapidity $k_j$
represents the quasi-momentum of a beam of charges,
while complex $k_j$'s, which
are sometimes called string solutions
because they appear in a linear group with a common real part,
represent bound states of electrons.

For a set of real $k_j$'s, we can take a logarithm of
Bethe-ansatz equations to have
\begin{equation}
\label{realk}
k_jN_a = 2\pi I_j + 2\pi\Phi/\Phi_0-\sum_{\alpha =1}^M
2 \tan^{-1}\frac{4}{U} (\sin k_j - \Lambda_\alpha) \; ,
\end{equation}
where $M$ is the number of down spins, $t$ the transfer, $U$
the Hubbard interaction, and
$I_j$ ($j=1,\ldots,N$) is an integer
(half odd integer) for an even (odd) $M$.
We assume that spin rapidities $\Lambda_{\alpha}$ are real
as is the case with the ground state of the repulsive model.
Equation (\ref{realk}) has a periodicity of $N_a$ in $I_j$.
Since an increase of $\Phi$ by $\Phi_0$ causes
a uniform shift of $I_j$ by unity,
a real solution indeed has a periodicity of $N_a\Phi_0$
(except at the half-filling, see below).

Now, the string solutions for charge rapidities,
$\{ k_{nl}$ ($l=1,\ldots,2n) \}$,
are specified by a single real parameter,
${\Lambda'}_{n}$\cite{takahashi}.
When all the $N$ particles form a single bound state
(an $N$-bound state in Sutherland's words),
the equation reduces to
\begin{eqnarray}
& \exp & (iN_a\sum_l k_{nl} ) = \nonumber \\
& \exp & \left\{
-iN_a[\sin^{-1}({\Lambda'}_n+inU/4)+\sin^{-1}({\Lambda'}_n-inU/4)]
\right\}
= \exp(i2\pi N\Phi/\Phi_0) \; .
\end{eqnarray}
We can then see that $\Lambda'_n$ is determined by
the total momentum $\sum_l k_{nl}$,
so that a change of $\Phi$ by $\Phi_0/N$ is enough to
shift ${\Lambda'}_n$ to the next position among the $N_a$ solutions.
Thus we end up with a period of $N_a\Phi_0/N$ for a single $N$-bound state.

Solutions having more than one set of strings
have been known to exist
for e.g., the attractive Hubbard model having a set of two-strings
(electron pairs)\cite{woyna,shulott}.
However, their high-energy spectra has not been fully understood,
so that the spectral flow has to be obtained numerically even in 1D.

So we move on to the numerical implementation
of looking at the flow, which is readily applicable to
two or higher spatial dimensions, and also to various models such as
the extended Hubbard or
$t$-$J$ models.  In determining the extended period,
we have to keep track of the ground
state for the range of the flux well beyond the flux quantum,
where the energy level soars to become high-energy states.
In addition the level has to be traced straight through level crossings,
so that the numerical method must be good enough to reproduce
(a) level crossings and (b) high-energy states.
Conventional methods such as the exact diagonalization or
quantum Monte Carlo methods
would then be inadequate.

The algorithm we propose consists of successive estimations of
the energy and wave function for $\Phi+\Delta \Phi$.
The new state $\Psi$ is estimated by multiplying a connection,
$e^{iA(\Phi)\Delta \Phi}$, defined as
\begin{equation}
\label{Bphase}
 \Psi(\Phi+\Delta \Phi)=e^{iA(\Phi)\Delta \Phi}\Psi(\Phi).
\end{equation}
Here a matrix $A$ is approximately diagonal
($e^{iA_i\Delta \Phi} \simeq \Psi_i(\Phi+\Delta \Phi)/\Psi_i(\Phi)$)
for a small enough $\Delta \Phi$,
where $\Psi_i$ denotes the $i$-th component.
Namely the connection reduces to
a set of pure phases (i.e., Berry's geometrical phase).
Each estimation is followed by a correction in the following recursive manner.

\noindent i) Start from the ground state $\Psi (\Phi =0)$
and the next one $\Psi (\Delta\Phi)$
which are obtained by the Lanczos method.

\noindent ii) Predict the connection from the previous step as
$e^{iA_i\Delta \Phi} \simeq \Psi_i(\Phi)/\Psi_i(\Phi-\Delta \Phi)$
to produce the next one,
$\tilde{\Psi}(\Phi+\Delta \Phi)=e^{iA(\Phi )\Delta \Phi}\Psi(\Phi)$.

\noindent iii) Correct $\tilde{\Psi}$ by operating some power, $m$, of the
resolvent as
\begin{equation}
\left\{ \frac{1}{[H(\Phi +\Delta\Phi )-E]^2} \right\}^m
\tilde{\Psi}(\Phi +\Delta\Phi)
\stackrel{m\rightarrow \infty}{\longrightarrow}
\Psi(\Phi +\Delta\Phi)
\label{iter}
\end{equation}
This power method is performed with the inverse-iteration
and the conjugate-gradient optimization.
$E$ in the resolvent can be calculated accurately
from the wavefunction updated every time we increase $m$.

\noindent iv) Check the convergence,
$|(H(\Phi +\Delta\Phi )-E)\Psi (\Phi +\Delta\Phi)|<\varepsilon$,
for a given accuracy $\varepsilon$.
Increase $m$ until this inequality is fulfilled,
then move on to ii).

Applied over the extended zone, the method can count
the periodicity (or the `winding number') of the state.
The tractable matrix size is the similar to
that for the conventional Lanczos diagonalization.

Around level crossings the calculation becomes subtle.
We first observe that level crossings come in two classes:
One class occurs for two levels that differ in some symmetry of the system.
The second occurs at a phase transition point,
{\it e.g.,} the normal-superconductor transition.
The first class can be dealt with by only letting
the convergence criterion in iv) with typically $\varepsilon=10^{-8}$
for the systems considered here
with the matrix size up to a few ten thousands.
The second class requires scaling analysis by varying
the distance from the critical point as illustrated
for the $t-J$ ladder below.

Our first example is the 1D Hubbard model.
Figure 1 compares the spectral flow for
a 10-site ring with 6 electrons for a repulsive model with $U/t=+1$ and
an attractive model with $U/t=-1$.
The extended period, which is $N_a\Phi_0$
for the repulsive model,
is indeed halved for the attractive case,
providing a clear-cut detection of the two-electron pairing.

For comparison, we have superposed the anomalous flux quantization test
for $0\leq \Phi <\Phi_0$ in Fig.1.
We can see that the latter test is indeed ambiguous.
To be more precise, we can observe the following.
If the Fermi sea has a closed shell ($N\equiv 2$ (mod 4), as is the case with
Fig.1), the true ground state is always spin singlet irrespective of
the sign of $U$.
In addition to the branch starting from the ground state, there is a
second branch, which is degenerate with the first one
at $\Phi=\Phi_0/2$ for $U=0$ but shifts upward
for a repulsive interaction, or downward
for an attractive interaction.
Thus a dip appears in a continuous fashion as a negative $U$ is turned on.
However, the half periodicity can appear even for the repulsive interactions
for open-shell Fermi seas\cite{ferretti}.
This is due to the existence of a spin-triplet state, which
is degenerate with the ground state
at $\Phi=0$ and stabilized over the singlet state for $U>0$\cite{ka}.
If we further go into the strong correlation limit, even more
anomalous $1/N_e$-periodicity can appear as shown
by Kusmartsev\cite{kusmar}.
Thus we have to worry about these finite-size effects
due to other branches lying around to obscure the period in
the anomalous flux quantization.
In two or higher dimensions
such situations may be improved\cite{scalapino,assaad},
but we should re-emphasize that
the winding-number counting here concentrates on the
adiabatic evolution of a {\it single} state, where
the `global' period is determined independently of
other branches.

The abrupt change from $N_a\Phi_0$ to $N_a\Phi_0/2$
in the extended AB period takes place exactly at $U=0$,
which is the critical point in 1D.
The change occurs despite the fact that
three spectral flows (for $U<0, U=0, U>0$)
are almost identical
around the respective minimum except for some offsets.
This corresponds to the known fact\cite{scalapino} that
the charge stiffness (or Drude weight) does not exhibit singular jumps
even at the critical point when the system is finite.

A closer inspection shows that the long AB period for $U\neq 0$ is
dominated by
some level crossings that turn into a level repulsion or
level {\it anti\,}crossings, where
different sets of anticrossings are selected according as $U>0$ or $U<0$.
It is at first puzzling how such a qualitative change can possibly
occur for an infinitesimal $|U|$,
since charge rapidity $k_j$'s coalesce into doubly occupied $k$ points
in the Fermi sea
no matter how the critical point ($U=0$) is approached from
repulsive or attractive sides (see Ref.\cite{shultz}).

The puzzle is resolved by looking at the weak-coupling Bethe-ansatz behavior.
As the $k$ points shift with $\Phi$,
the crucial level anticrossing occurs
when the upper-most doubly-occupied $k$ point reaches
$k_0=\pi/2$ situated at $\varepsilon (k_0)=0$,
the center of the band dispersion.
This special point accommodates highly degenerate
states that are connected by
two-particle scattering processes that comprise the normal ones,
$c_{k_0 \uparrow}^\dagger c_{k_0 \downarrow}^\dagger
\rightarrow
c_{k_0+p \uparrow}^\dagger c_{k_0-p \downarrow}^\dagger$,
with $p$ the momentum transfer, along with
the Umklapp process,
$c_{k_0 \uparrow}^\dagger c_{k_0 \downarrow}^\dagger
\rightarrow
c_{k-2k_0 \uparrow}^\dagger c_{k-2k_0 \downarrow}^\dagger$.
When $U$ is switched on, the degenerate perturbation
dictates that some of these states must mix to give anticrossings.

The normal process with $p=2\pi N/N_a$ selectively
produces the anticrossing for the repulsive case.
In contrast, the anticrossing in the attractive case
is caused by the Umklapp process that transfers a pair
to other $\varepsilon =0$ points, which
is the only possible process indeed if the pairs are not dissociated
by the adiabatic change in $\Phi$.
The fact that $k_0$ is the key position is illustrated in Fig.1:
it takes $\Phi_c/\Phi_0 = (k_0-k_F)/\Delta k$ for
the upper-most doubly occupied
$k$ point ($k_F$) to reach $k_0$,
where $\Phi_c/\Phi_0 = 1.5$ for $N_e=6, N=10$, in exact
agreement with Fig.1.

Given this property, we can readily
show that a series of dissociations of
doubly occupied states for $U>0$ gives the extended period of
$N_a\Phi_0$, while a series of
Umklapped doubly occupied states for $U<0$ halves it.
Thus, an infinitesimal interaction is enough to change
the global topology (the winding number)
of the connection in this example in 1D.

A fuller understanding in terms of the spin-charge separation in 1D
emerges if we look more closely at the Bethe ansatz,
where $\Lambda_{\alpha}$ (a `spin degrees of freedom'),
not directly coupled with $\Phi$, stays approximately constant,
while the `charge degrees of freedom' $k_i$ progressively
changes with $\Phi$ for a repulsive $U$.
This is accomplished by $\Lambda_{\alpha}$
sequentially parting company with
one $k_i$ to meet another, which is exactly where the anticrossings occur.
On the other hand, each $\Lambda_{\alpha}$ is attached to a pair,
$k_j, k_j^*$, for an attractive $U$, and
the center-of-mass momentum, 2Re($k_j$), of the pair remains
within the first Brillouin zone due to the Umklapp process,
or equivalently, $k_j$'s have to satisfy cos$(k_j)\geq 0$
for the two-bound solutions to exist.
The details will be published elsewhere.

In order to demonstrate that the present method is applicable to
the case in which the Bethe-ansatz is inapplicable, and also
to provide a step toward higher dimensions,
we move on to the $t$-$J$ ladder model.
This model, originally conceived for some copper or vanadium oxides,
is believed to exhibit superconductivity for small hole doping
from the half-filling\cite{t-Jladder}.
Here we perform the extended AB test for
a $6\times 2$-site system with 4 electrons,
which corresponds to a
hole concentration of $n=2/3$.

In the result, Fig.2, we can clearly see that
the periodicity is halved into $N_a/2$ as $J$ is increased to $J>2.04t$,
which indicates paring.
Here we can illustrate a nice feature about the extended AB test:
since the transition is associated
with a level crossing (a cusp) turning into an anticrossing,
we can numerically plot (inset of Fig.2)
the size, $\Delta$, of the level repulsion against
the relevant parameter ($J$ here) to identify the critical point
($J_c\simeq 2.04$ here) at which $\Delta$ vanishes.
Thus we can estimate the critical value $J_c$ for a finite system
in a well-defined manner, which may then be cast into a
finite-size scaling analysis for a more pertinent definition of the
critical point.
Thus the present method provides a possible way to
determine the phase diagram of a given model.

So far we have not discussed anything directly about the
coherence of the pairs, so that we are talking about a necessary
condition for superconductivity.
For that matter, Byers-Yang theorem also
gives a necessary condition for the superconductivity (or
the Meissner effect).
It is an intriguing future problem to see if the coherence can
possibly appear in the spectral flow.

Another comment is that the present test gives information on
metal-insulator transitions such as the Mott-Hubbard transition as well.
For the half-filled Hubbard model (a Mott's insulator),
the extended AB period reduces down to $\Phi_0$.
The sudden change from the full period $N_a\Phi_0$ to the
minimum one is due to an appearance of the charge gap,
across which the flow is inhibited to jump so that
the system has to return to the original state as soon as the flux
reaches $\Phi_0$, in analogy with the gauge argument
of the quantum Hall effect.
Thus we can expect that the present method can detect
the existence or otherwise of the Fermi surface\cite{ss2}.
A more interesting example is the phase separation, where
the system should respond as a single bound state as will be
reported elsewhere.
The different winding numbers for paired, metallic and insulating phases
may be analyzed in terms of the homotopy of the phase-space
(or the fiber bundle) of correlated electron systems.

The authors wish to thank
Kazuhiko Kuroki and Fakher Assaad for valuable discussions.
Numerical calculations were performed on FACOM VPP500 in Supercomputer Center,
Institute for Solid State Physics, University of Tokyo.
This work was supported in part by a Grant-in-Aid from
the Ministry of Education, Science and Culture, Japan.

\begin{figure}
\caption{
The result for the spectral flow for
a 10-site Hubbard ring with 6 electrons.
The flow lines represent a repulsive case ($U=t$, solid line),
an attractive case ($U=-t$, dashed line), and the non-interacting case
(dotted line).
The relevant level crossing point is marked with $\Phi_c$.
For comparison, the low-lying flows over one period of the flux quantum
is reproduced.
}
\end{figure}

\begin{figure}
\caption{
The spectral flow lines for
a $t$-$J$ ladder model
with 4 electrons in $6\times2$ sites
for $J=2.2, 2.6, \ldots , 3.8$
>from top to bottom.
The inset depicts the size, $\Delta$,
of the relevant level repulsion against $J$, where the
line is a guide to the eye.
}
\end{figure}

\end{document}